\documentclass[sigconf,natbib]{acmart}
\usepackage{multirow}
\usepackage{makecell}
\usepackage{lipsum}
\usepackage{subcaption}
\usepackage{float}

\renewcommand\footnotetextcopyrightpermission[1]{}

\AtBeginDocument{%
  \providecommand\BibTeX{{%
    \normalfont B\kern-0.5em{\scshape i\kern-0.25em b}\kern-0.8em\TeX}}}

\begin{document}

\title{CaseEncoder: A Knowledge-enhanced Pre-trained Model for Legal Case Encoding}

\author{Yixiao Ma}
\email{mayx20@mails.tsinghua.edu.cn}
\affiliation{%
  \institution{Department of Computer Science and Technology, Institute for Internet Judiciary, Tsinghua University. Quan Cheng Laboratory}
  \city{Beijing}
  \country{China}
  \postcode{100084}
}

\author{Yueyue Wu}
\email{wuyueyue1600@gmail.com}
\affiliation{%
  \institution{Department of Computer Science and Technology, Institute for Internet Judiciary, Tsinghua University. Quan Cheng Laboratory}
  \city{Beijing}
  \country{China}
  \postcode{100084}
}

\author{Weihang Su}
\email{swh22@mails.tsinghua.edu.cn}
\affiliation{%
  \institution{Department of Computer Science and Technology, Institute for Internet Judiciary, Tsinghua University. Quan Cheng Laboratory}
  \city{Beijing}
  \country{China}
  \postcode{100084}
}

\author{Qingyao Ai}
\email{aiqingyao@gmail.com}
\affiliation{%
 \institution{Department of Computer Science and Technology, Institute for Internet Judiciary, Tsinghua University. Quan Cheng Laboratory}
  \city{Beijing}
  \country{China}
  \postcode{100084}
}

\author{Yiqun Liu}
\authornote{Corresponding author}
\email{yiqunliu@tsinghua.edu.cn}
\affiliation{%
  \institution{Department of Computer Science and Technology, Institute for Internet Judiciary, Tsinghua University. Quan Cheng Laboratory}
  \city{Beijing}
  \country{China}
  \postcode{100084}
}

\renewcommand{\shortauthors}{Ma et al.}

\begin{abstract}

Legal case retrieval is a critical process for modern legal information systems. While recent studies have utilized pre-trained language models (PLMs) based on the general domain self-supervised pre-training paradigm to build models for legal case retrieval, there are limitations in using general domain PLMs as backbones. Specifically, these models may not fully capture the underlying legal features in legal case documents. To address this issue, we propose CaseEncoder, a legal document encoder that leverages fine-grained legal knowledge in both the data sampling and pre-training phases. In the data sampling phase, we enhance the quality of the training data by utilizing fine-grained law article information to guide the selection of positive and negative examples. In the pre-training phase, we design legal-specific pre-training tasks that align with the judging criteria of relevant legal cases. Based on these tasks, we introduce an innovative loss function called \textit{Biased Circle Loss} to enhance the model's ability to recognize case relevance in fine grains. Experimental results on multiple benchmarks demonstrate that CaseEncoder significantly outperforms both existing general pre-training models and legal-specific pre-training models in zero-shot legal case retrieval. The source code of CaseEncoder will be released when the paper is published.

\end{abstract}


\maketitle

\section{Introduction}

Legal case retrieval is a critical process for modern legal information systems, as it aims to find relevant prior cases (i.e., precedents) that serve as important references to the case to be judged. In recent years, pre-trained language modeling (PLM) techniques have achieved great success in general-domain retrieval tasks, which has also led to attempts to apply PLMs in the legal domain. For example, \citet{zhong2019openclap} and \citet{xiao2021lawformer} propose a legal-oriented PLM based on BERT~\cite{devlin2018bert} and Longformer~\cite{beltagy2020longformer}, respectively. However, existing legal-oriented PLMs have limitations in their adaptation to the legal domain, because they mainly rely on replacing general-domain training data with legal data or extending the input length to fit the long-length characteristic of legal documents. \citet{ge2021learning} parse articles in the form of premise-conclusion pairs to train a multi-level matching network for legal case matching. \citet{bhattacharya2022legal} propose Hier-SPCNet which substantially improves the network-based similarity by introducing legal textual information. However, they still adopt general-domain PLMs as the foundation of their approach. While interpretable, such models do not fully enable the PLM to comprehend legal concepts in case documents, which limits their generalizability and applicability.



To address this issue, this paper proposes CaseEncoder, a PLM that leverages fine-grained legal knowledge to improve both the data sampling phase and the pre-training phase. In the data sampling phase, we split law articles into unambiguous branches and construct inner logical relations in each article, which are used to estimate relevance weights between cases. These relevance weights serve as pseudo labels to guide the selection of positive and negative cases. In this way, our proposed data sampling method improves the quality of sampled cases for further pre-training. In the pre-training phase, CaseEncoder adopts masked language modeling (MLM) and fine-grained contrastive learning tasks. These tasks aim to match two main concepts in the judging criteria of legal relevance: \textit{key circumstances} and \textit{key elements}, respectively. \textit{Key circumstances} refer to significant case descriptions in the document, while \textit{key elements} are the legal-level abstraction of \textit{key circumstances}, which focus more on consistency with law articles. In practice, the MLM task captures the semantic-level case descriptions, while the fine-grained contrastive learning task adopts sampled cases together with their relevance weights to enhance the model's ability to identify \textit{key elements}. Based on the design of the fine-grained contrastive learning task, we propose an innovative loss function, Biased Circle Loss, which leverages the obtained fine-grained relevance score to optimize the recognition of \textit{key elements}.

Experimental results on multiple legal case retrieval datasets demonstrate that CaseEncoder significantly outperforms the existing general pre-training models and legal-specific pre-training models. We also present case document embedding visualizations to showcase the potential of CaseEncoder in downstream tasks such as charge prediction and article prediction. The source code of CaseEncoder will be released when the paper is published.

\section{Task Definition}

Given a query case $q$, the task aims to retrieve relevant cases from a candidate list $L = \{ c_1, c_2, ..., c_M \}$, where $M$ is the size of $L$, and rank them by the relevance to $q$. Each candidate case document in the list has three main components: 1. \textit{Facts} are objective fact statements confirmed by the court based on the evidence provided by the defendant and plaintiff. These statements typically answer questions such as where, when, and how the case occurred. 2. \textit{Holding} is the judge's opinion on the key arguments of the case. It explains the reasoning behind the judge's decision. 3. \textit{Decision} contains the final judgment of the defendant including the charge, sentence, and articles involved. This component is the official outcome of a case.

In most legal case retrieval scenarios, $q$ only contains the \textit{Facts} component, while each candidate case includes title, meta information, \textit{Facts}, \textit{Holding}, \textit{Decision}, and related law articles. In this paper, we focus primarily on retrieving cases under criminal law.

\section{Method}

\begin{figure}[ht]
    \centering
    \includegraphics[width=\linewidth]{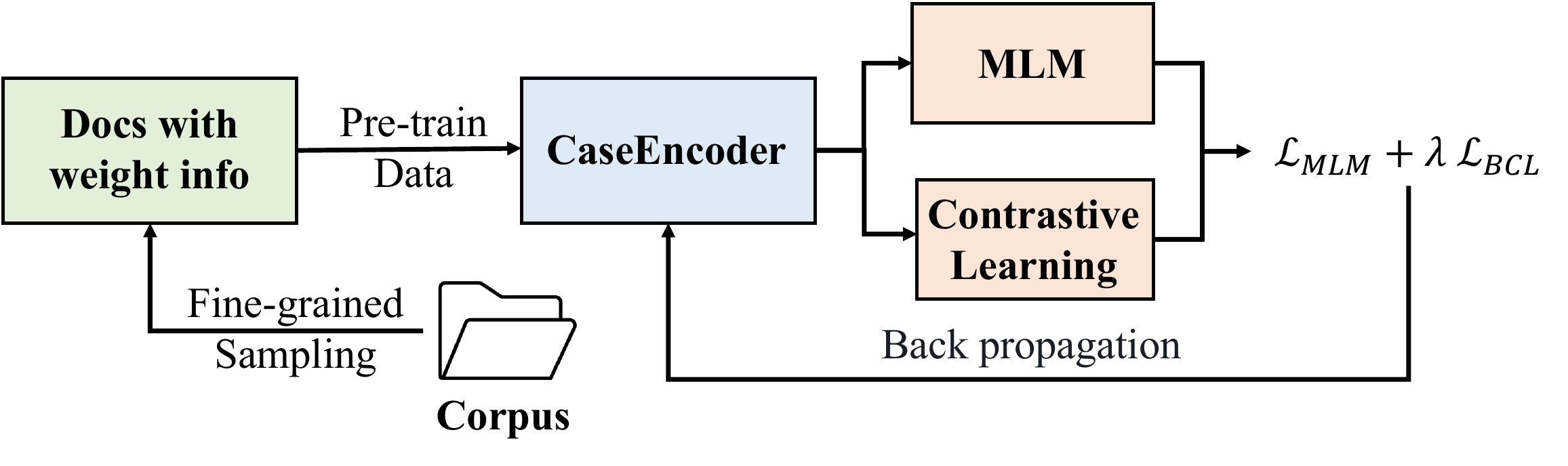}
    \caption{The overall framework of CaseEncoder.}
    \label{fig:overall}
    \vspace{-3mm}
\end{figure}

This section outlines the design and implementation of CaseEncoder. Figure~\ref{fig:overall} illustrates the overall framework of the model. We begin by introducing the fine-grained case sampling method used for data preparation in Section~\ref{sec:fg sample}. Then, in Section~\ref{sec:pre-train task}, we describe the pre-training tasks proposed in CaseEncoder.

\subsection{Fine-grained Case Sampling}
\label{sec:fg sample}

\begin{figure*}[ht]
    \vspace{-4mm}
    \centering
    \includegraphics[width=\linewidth]{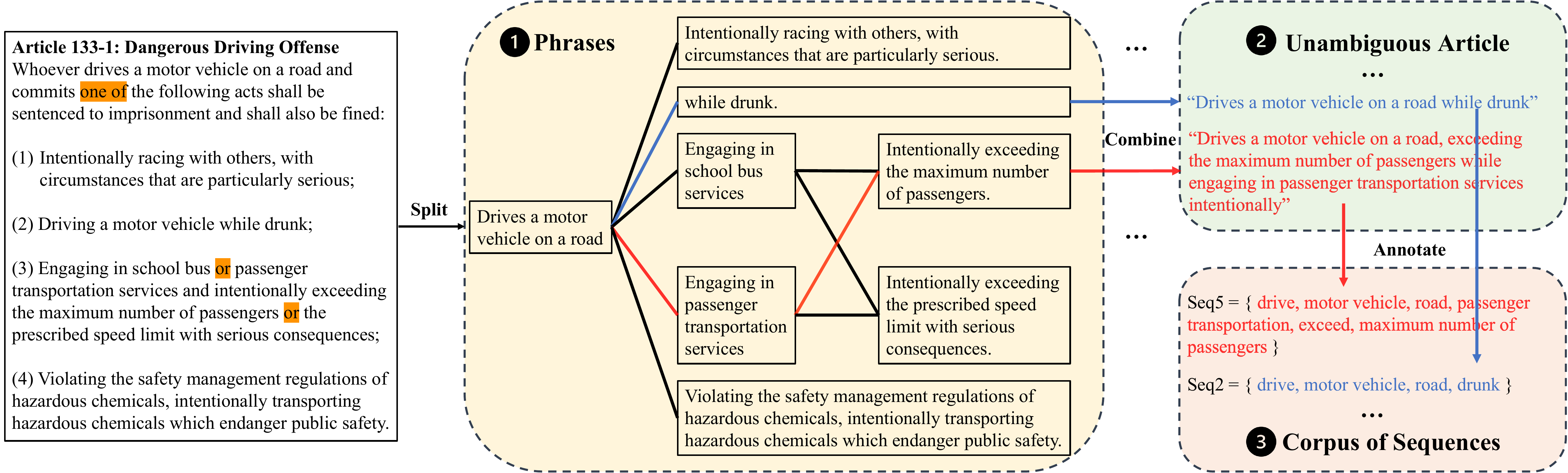}
    \caption{An illustration of the process of collecting a fine-grained unambiguous article corpus.}
    \label{fig:article}
    \vspace{-4mm}
\end{figure*}

Recent studies on legal-oriented PLMs do not focus on understanding legal texts from a bottom-up approach. In other words, most PLMs simply replace the general-domain training corpus with legal texts without considering the legal correlation between these texts. This is mainly because the annotation of legal cases is time-consuming and requires much expertise, making it challenging to collect large-scale labeled data. On the other hand, in contrastive learning, which has proven to be effective in the pre-training phase, data needs to be sampled in advance as positive and negative cases. Unlike the general domain, in the legal domain, it is not appropriate enough to sample the positive and negative legal cases simply based on raw information in documents (e.g., charges, law articles, etc.). Because in a real legal scenario, a judge will decide a case based on how well \textit{key circumstances } and \textit{key elements} of the case match the constituent elements (fine-grained interpretation of the law article). Therefore, this paper proposes a fine-grained sampling method for legal case documents with reference to the specific process of judges deciding cases. By doing so, the sampled positive and negative cases can match the manually labeled relevance as much as possible for the subsequent contrastive learning task.

Generally, a law article covers multiple branches. Each branch describes a certain situation applicable to this article. For example, the article shown in Figure~\ref{fig:article} has seven branches in total. In the third act of this article, 'engaging in school bus service' and 'engaging in passenger transportation service' belongs to different branches even if they appear in the same sentence. That is to say, phrases in one article have not only a sequential relationship but also a parallel relationship. The above example phrases belong to a parallel relationship. Phrases are combined in a permutation way to generate branches without ambiguity. We call the branches of the article as \textit{unambiguous articles}. Any case that matches any of the unambiguous articles belongs to this article. Obviously, cases under the same article may belong to different unambiguous articles. Therefore, the relevance between two cases can not be directly determined by reason like 'they all belong to Article 133-1'. In this paper, we aim to identify more fine-grained article information (i.e., unambiguous articles) for each case as the preliminary of finding positive and negative cases in our case sampling algorithm. To this end, we divide all law articles into fine-grained information similar to Figure~\ref{fig:article}: First, we split an original article into unambiguous articles. Then, legal keywords are manually annotated in each unambiguous article, which is under the guidance of legal experts. Finally, we extract the annotated keywords to form a word-level sequence representing the corresponding unambiguous article. All such sequences constitute an unambiguous article corpus $C_{a} = \{seq_1, seq_2, ..., seq_T \}$, where $T \in \mathbb{N}^+$ is the number of sequences extracted from one article.

Then, for cases committing the same crime, our case sampling strategy can recognize their relevance to each other with the term-level similarity to the sequences in $C_{a}$. Specifically, given a case, the first step is to extract its \textit{Holding} from the case document, because \textit{Holding} contains reasons for the final judgment, which is highly related to the articles of the case. Let the extracted \textit{Holding} is denoted as $h$, we can compute a similarity vector by:
\begin{gather}
    v = [ \operatorname{BM25}(seq_1, h), \operatorname{BM25}(seq_2, h), ..., \operatorname{BM25}(seq_T, h) ] \in \mathbb{R}^T
\end{gather}
where $\operatorname{BM25}$ is the traditional BM25~\cite{robertson1995okapi} model initialized with corpus $C_{a}$, and $\operatorname{BM25}(seq_i, h)$ denotes the BM25 score between the keyword sequence of the i-th unambiguous article and $h$. $v$ can be intuitively interpreted as the feature of a case at the law article level. Next, given two cases $c_i$ and $c_j$, the legal-specific relevance weight $w_{ij}$ can be presented as:
\begin{gather}
    w_{ij} = \frac{A_{i} \cap A_{j}}{|A_i|} \times \operatorname{rel}(c_i, c_j) \label{eq:w}\\ 
    \operatorname{rel}(c_i, c_j) = \begin{cases}
        1 & \operatorname{argmax}(v_i) = \operatorname{argmax}(v_j) \\
        \operatorname{max}(\operatorname{cos}(v_{ik}, v_{jk})) & Otherwise
    \end{cases} \label{eq:rel}
\end{gather}
where $A_{i}$ is the set of articles involved in case $c_i$, $\operatorname{cos}$ is the cosine similarity score, and $k$ is the k-th article in $A_{i}$. In other words, the relevance weight between two cases is mainly determined by the extent to which their related law articles overlap. If two cases, $c_i$ and $c_j$, are both most similar to the same unambiguous article, then they are considered the most relevant cases. Otherwise, the relevance weight will be decayed, depending on the cosine similarity between their similarity vectors $v_i$ and $v_j$. Note that a maximum function is added to Equation~\ref{eq:rel}, as $\operatorname{rel}(c_i, c_j)$ is valued by the most relevant scenario of all $A_{i} \cap A_{j}$. 

Finally, for each case $c_{i}$ in the legal case corpus, we can sample a legally explainable positive case $c_{i+}$ according to $w$. The input data for the pre-training phase will be in the form of quadruples $(c_i, c_{i+}, v_i, v_{i+})$.

\subsection{Legal-specific Pre-training Task}
\label{sec:pre-train task}

As a law-oriented pre-trained model, we aim to integrate legal knowledge into the design of pre-training tasks, so that the model can acquire the ability to understand case documents not only at the semantic level but also at the legal concept level after training. To this end, in this paper, we refer to the judging criteria of relevant cases to design our pre-training tasks. As demonstrated by \citet{ma2021lecard}, two cases are relevant if they satisfy two requirements: high similarity between their \textit{key circumstances}, and high similarity between their \textit{key elements}. Specifically, \textit{key circumstances} refer to significant case descriptions, while \textit{key elements} focus more on the consistency with law articles and represent the legal-level abstraction of \textit{key circumstances}. In summary, a case is considered relevant to another when the case description and abstracted legal concept are both relevant.

Following the idea of the judging criteria, two pre-training tasks are adopted in this paper. The first pre-training task is the masked language modeling (MLM) task, which enables the capture of regular semantic-level meaning of case description. As discussed in \citet{devlin2018bert, liu2019roberta, ma2021prop}, MLM contributes to producing embeddings with contextual information. Such embeddings are beneficial to the representation of \textit{key circumstances} in legal cases. In detail, we only select \textit{Facts} in a case document to randomly mask 15\% of the tokens for MLM, because \textit{key circumstances} are all included in \textit{Facts}. The masked text is then fed into CaseEncoder to predict the masked tokens based on the surrounding unmasked tokens. The MLM loss function is defined as:
\begin{gather}
    \mathcal{L}_{MLM}=-\sum_{x^{\prime} \in m(\mathbf{x})} \log p\left(x^{\prime} \mid \mathbf{x}_{\backslash m(\mathbf{x})}\right)
\end{gather}
where $\mathbf{x}$ is the text in \textit{Facts}, $m(\mathbf{x})$ is the set of masked tokens, and $\mathbf{x}_{\backslash m(\mathbf{x})}$ is the set of unmasked tokens.

The second pre-training task is a fine-grained contrastive learning task that utilizes the information from quadruples $(c_i, c_{i+}, v_i, v_{i+})$ obtained in Section~\ref{sec:fg sample}. The contrastive learning task in previous work~\cite{chen2020simple} train a model using augmented positive cases and regards the rest of the cases in the same batch as negatives. However, in the legal domain, the relevance scale is more fine-grained. One legal case can be partially relevant to another, and the extent of relevance is mostly determined by the previously mentioned \textit{key elements}. Therefore, a fine-grained contrastive learning task is proposed to enhance the recognition of \textit{key elements}. Specifically, suppose the batch size is $N$ and each quadruple has two cases, the total number of cases in a batch is $2N$. First, a multi-layer Transformer is adopted to obtain the representations of $2N$ cases. Then, we take the output of $[\operatorname{CLS}]$ token in the last hidden layer of Transformer as the case embedding: $e_1, e_2, ..., e_{2N}, e_i \in \mathbb{R}^H$, where $H$ is the hidden size. Finally, the training objective of this fine-grained contrastive learning task, Biased Circle Loss (BCL), is defined as:
\begin{gather}
        \mathcal{L}_{\text {BCL}}=\log [1+\sum_{j=1}^L \exp (\gamma \alpha_n^j(s_n^j-\Delta_n)) \sum_{i=1}^K \exp (-\gamma \alpha_p^i(s_p^i-\Delta_p))] \label{eq:bcl}\\ 
        \alpha_p^i=|\exp^{w_p - 1} \cdot O_p-s_p^i|, \alpha_n^j=[s_n^j-O_n]_{+}
\end{gather}
where $s_p$ and $s_n$ are cosine similarity scores of between-class and within-class case embeddings, respectively. $\alpha_p$ and $\alpha_n$ are parameters controlling the speed of convergence, where $\alpha_p$ is determined by the legal-specific relevance weight in Equation~\ref{eq:w}. $\gamma$, $O_p$, $O_n$, $\Delta_p$, and $\Delta_n$ are hyper-parameters of scale factor, optimum for $s_p$, optimum for $s_n$, between-class margin, and within-class margin, respectively. In this way, CaseEncoder is trained to pull case embeddings in the same class closer and push case embeddings in different classes apart. The distance between case embeddings in the vector space depends on the value of $s_p$ and $s_n$.

There are two main differences between our proposed loss function and Circle Loss~\cite{sun2020circle}: First, we expand the original loss function from a binary scenario to a multi-class scenario since legal cases within a batch can be classified into multiple classes. In detail, we consider any two cases to be in the same class if their legal-specific relevance weight is larger than a particular threshold $W_T$, and such a rule is transitive across all cases in a batch. Therefore, the actual implementation of Equation~\ref{eq:bcl} is more complicated because cases are of multiple classes. Second, we add a weight parameter $\alpha$ to $\mathcal{L}_{\text {BCL}}$ to account for the extent of relevance between cases. By taking the legal-specific relevance weight into consideration, CaseEncoder is trained to discriminate between relevant cases in fine grains.


Finally, CaseEncoder is optimized by the linear combination of MLM loss and BCL loss, where $\lambda$ is a hyper-parameter:
\begin{gather}
    \mathcal{L}_{\text {total}} = \mathcal{L}_{\text {MLM}} + \lambda\mathcal{L}_{\text {BCL}}
\end{gather}

\section{Experiments}

\begin{table*}[]
\small
\vspace{-4mm}
\setlength{\abovecaptionskip}{-0.1mm}
\caption{Evaluation results of different models on LeCaRD, CAIL2021-LCR, and CAIL2022-LCR. $\dag \backslash \ddag \backslash \S$ denote statistical significance compared to CaseEncoder at a level of p-value $< 0.05 \backslash 0.01 \backslash 0.005$ using Wilcoxon test.}
\centering
\begin{tabular}{l|lll|lll|lll}
\toprule
& \multicolumn{3}{c|}{LeCaRD} & \multicolumn{3}{c|}{CAIL2021-LCR} & \multicolumn{3}{c}{CAIL2022-LCR} \\
\multicolumn{1}{c|}{\textbf{Model}} & \multicolumn{3}{c|}{\textbf{NDCG@(10, 20, 30)}} & \multicolumn{3}{c|}{\textbf{NDCG@(10, 20, 30)}} & \multicolumn{3}{c}{\textbf{NDCG@(10, 20, 30)}} \\ \hline
\multicolumn{1}{c|}{BERT-XS} & \multicolumn{1}{c}{0.343$^\S$} & \multicolumn{1}{c}{0.355$^\S$} & \multicolumn{1}{c|}{0.384$^\S$} & \multicolumn{1}{c}{0.361$^\S$} & \multicolumn{1}{c}{0.368$^\S$} & \multicolumn{1}{c|}{0.388$^\S$} & \multicolumn{1}{c}{0.358$^\S$} & \multicolumn{1}{c}{0.359$^\S$} & \multicolumn{1}{c}{0.383$^\S$} \\
\multicolumn{1}{c|}{Lawformer} & \multicolumn{1}{c}{0.620$^\S$} & \multicolumn{1}{c}{0.623$^\S$} & \multicolumn{1}{c|}{0.636$^\S$} & \multicolumn{1}{c}{0.691$^\S$} & \multicolumn{1}{c}{0.684$^\S$} & \multicolumn{1}{c|}{0.699$^\S$} & \multicolumn{1}{c}{0.694$^\S$} & \multicolumn{1}{c}{0.688$^\S$} & \multicolumn{1}{c}{0.700$^\S$} \\
\multicolumn{1}{c|}{RoBERTa} & \multicolumn{1}{c}{0.748$^\dag$} & \multicolumn{1}{c}{0.762$^\ddag$} & \multicolumn{1}{c|}{0.791$^\S$} & \multicolumn{1}{c}{0.804$^\dag$} & \multicolumn{1}{c}{0.817$^\ddag$} & \multicolumn{1}{c|}{0.850$^\dag$} & \multicolumn{1}{c}{0.793$^\ddag$} & \multicolumn{1}{c}{0.803$^\S$} & \multicolumn{1}{c}{0.837$^\dag$} \\
\multicolumn{1}{c|}{RoBERTa-Legal} & \multicolumn{1}{c}{0.742$^\ddag$} & \multicolumn{1}{c}{0.764$^\S$} & \multicolumn{1}{c|}{0.806$^\S$} & \multicolumn{1}{c}{0.814$^\dag$} & \multicolumn{1}{c}{0.823$^\dag$} & \multicolumn{1}{c|}{0.855} & \multicolumn{1}{c}{0.800$^\ddag$} & \multicolumn{1}{c}{0.811$^\S$} & \multicolumn{1}{c}{0.846$^\dag$} \\
 \hline
\multicolumn{1}{c|}{CaseEncoder} & \multicolumn{1}{c}{\textbf{0.785}} & \multicolumn{1}{c}{\textbf{0.803}} & \multicolumn{1}{c|}{\textbf{0.839}} & \multicolumn{1}{c}{\textbf{0.842}} & \multicolumn{1}{c}{\textbf{0.849}} & \multicolumn{1}{c|}{\textbf{0.876}} & \multicolumn{1}{c}{\textbf{0.833}} & \multicolumn{1}{c}{\textbf{0.840}} & \multicolumn{1}{c}{\textbf{0.867}} 
\\ \bottomrule
\end{tabular}
\label{tab:overall}
\vspace{-4mm}
\end{table*}

\subsection{Experimental Settings}
\textbf{Evaluation.} As a retrieval task, all models in this paper are evaluated in three metrics: NDCG@10, NDCG@20, and NDCG@30. Additionally, to match the real legal scenario that the query is a case to be judged without any related information, all models are evaluated under the zero-shot setting. For a fair comparison, all models in this paper adopt a commonly used dual-encoder paradigm to retrieve legal cases. That is, both queries and candidate cases generate document-level embeddings by the same pre-trained encoder, and the final retrieved ranking list is sorted by the cosine similarity between embeddings.

\noindent
\textbf{Datasets and baselines.}
This paper adopts three publicly available datasets: LeCaRD, CAIL2021-LCR, and CAIL2022-LCR. LeCaRD~\cite{ma2021lecard} is the first Chinese legal case retrieval benchmark, which is widely used for the evaluation of retrieval models. CAIL2021-LCR~\footnote{\url{http://cail.cipsc.org.cn/task3.html?raceID=3&cail_tag=2022}} and CAIL2022-LCR~\footnote{\url{http://cail.cipsc.org.cn/task_summit.html?raceID=1&cail_tag=2021}} are two competition datasets. Since the experiment setting is zero-shot, both the training set and test set are included for the evaluation in this paper. CaseEncoder is compared to four PLMs: BERT-XS~\cite{zhong2019openclap}, Lawformer~\cite{xiao2021lawformer}, RoBERTa~\cite{liu2019roberta}, and RoBERTa-Legal. RoBERTa-Legal is a legal version of RoBERTa that conducts a secondary pre-training on legal data using the MLM task.

\noindent
\textbf{Parameter Settings.}
In this paper, all PLM backbones are imported from Huggingface~\cite{wolf2019huggingface}, with the learning rate set to $1*10^{-5}$. BM25 algorithm is implemented by Gensim~\cite{vrehuuvrek2011gensim}. The hyper-parameter for CaseEncoder is: $\gamma=16$, $O_p=1.25$, $O_n=0.25$, $\delta_p=0.75$, $\delta_n=0.25$, $W_T=0.25$, and $\lambda=\exp * 10^{-6}$. All training and experiments are conducted on eight 32G NVIDIA V100 GPUs. 

\vspace{-3mm}

\subsection{Experimental Results}
\label{subsec:experiment}

The overall results are shown in Table~\ref{tab:overall}. CaseEncoder outperforms baselines in terms of all metrics on three datasets, and most of the improvement is statistically significant. RoBERTa-Legal has the second-best overall performance and outperforms the original RoBERTa, which proves the idea in \citet{gururangan2020don} that a secondary pre-training using domain-specific data is beneficial to the overall performance in the target domain. Lawformer is not as effective as reported in \citet{xiao2021lawformer}. One possible explanation is that the case document adopted in this paper is relatively short, while Lawformer is trained specifically for long documents (4096 tokens). Besides, the effectiveness of BERT-XS~\cite{zhong2019openclap} is limited, because it utilizes Next Sentence Prediction (NSP) task for pre-training and its [CLS] token is not trained to represent document-level embeddings. These results demonstrate that our proposed CaseEncoder is effective in the retrieval task.

To investigate the effectiveness of our proposed fine-grained sampling method and contrastive learning task, we further conduct a series of ablation studies. As shown in Table~\ref{tab:abla}, removing the sampling method, loss function, or the contrastive learning task all lead to performance decline. Therefore, all of these innovations contribute to the effectiveness of CaseEncoder. In addition, Table~\ref{tab:abla} also indicates that BCL contributes most to the improvement of CaseEncoder, while the effect of adding a traditional binary contrastive learning task is limited.

\begin{table}[]
\small
\setlength{\abovecaptionskip}{-0.1mm}
\caption{Ablation study on LeCaRD. 'w/o sampling' denotes CaseEncoder without the fine-grained sampling method. 'w/o BCL' denotes replacing BCL with Circle Loss. 'w/o task' denotes CaseEncoder only pre-trained by MLM task.}
\centering
\begin{tabular}{c|ccc}
\toprule
\textbf{Model} & \textbf{NDCG@10} & \textbf{NDCG@20} & \textbf{NDCG@30} \\ \hline
CaseEncoder & \textbf{0.785} & \textbf{0.803} & \textbf{0.839} \\ \hline
w/o sampling & 0.778 & 0.795 & 0.827 \\
w/o BCL & 0.741 & 0.764 & 0.805 \\ 
w/o task & 0.742 & 0.764 & 0.806 \\\bottomrule
\end{tabular}
\label{tab:abla}
\vspace{-3mm}
\end{table}

\subsection{Application in Downstream Tasks}

\begin{figure}[]
    \centering
    \includegraphics[width=0.9\linewidth]{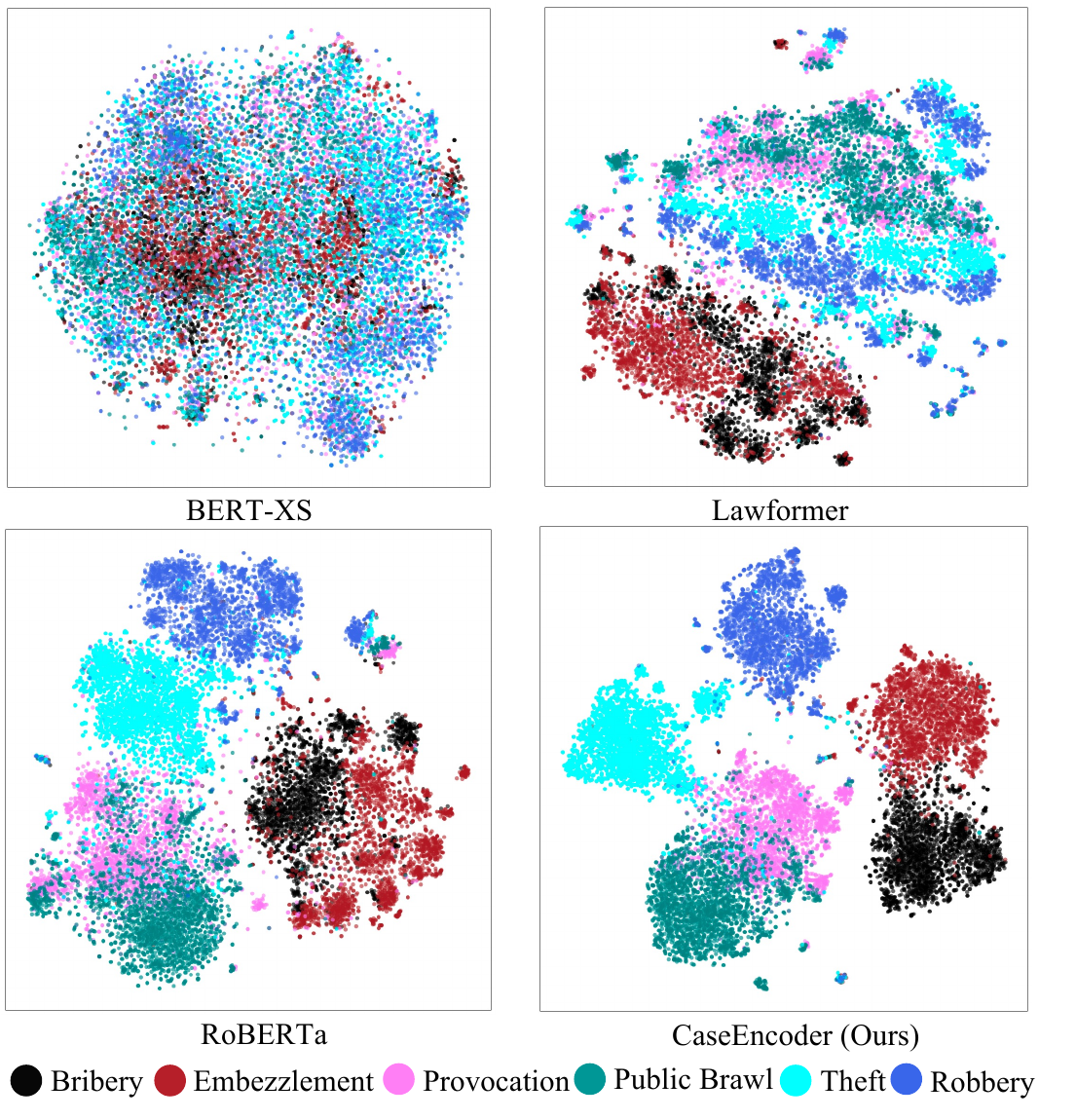}
    \caption{The visualization of case embeddings generated by four PLMs in the zero-shot manner.}
    \label{fig:visual}
    \vspace{-3mm}
\end{figure}

CaseEncoder is designed to effectively model case documents in the legal domain. In addition to the retrieval task, the document-level case embedding can also be utilized in other downstream tasks such as charge prediction. Figure~\ref{fig:visual} is an example of how the CaseEncoder improves the quality of case embeddings for charge prediction. We randomly select 2500 cases from each criminal charge and generate their corresponding case embeddings. Then, we use t-SNE~\cite{van2008visualizing} to reduce the dimension of case embeddings for visualization. Among all PLMs, CaseEncoder has the best ability to divide case embeddings into six clusters based on their charges, with only one pair of similar charges (Provocation and Public Brawl) having some overlap. By comparison, RoBERTa partially distinguishes between six charges, but with more overlap than CaseEncoder. The performance of BERT-XS and Lawformer is limited, which is consistent with the retrieval result and explanation in Section~\ref{subsec:experiment}. These visualizations demonstrate how the fine-grained legal knowledge embedded in CaseEncoder can be leveraged for a range of legal applications beyond case retrieval.

\section{Conclusion}

This paper proposes CaseEncoder, a pre-trained encoder that utilizes fine-grained legal knowledge to enhance the representation of case document embeddings. Experiments and visual analysis demonstrate the effectiveness of case embeddings generated by CaseEncoder in zero-shot legal case retrieval and other downstream legal tasks such as charge prediction.

\bibliographystyle{ACM-Reference-Format}
\bibliography{ref}

\end{document}